\documentclass[preprint2]{aastex}
\usepackage{epsf}
\usepackage{color}
\usepackage{graphicx}

\def\etal{{\it et al. }}
\def\ie{{\it i.e.,} }
\def\tg{t_{\rm g}}
\def\rg{r_{\rm g}}
\def\mbh{M_{\rm BH}}
\def\rinj{r_{\rm inj}}
\def\linj{l_{\rm inj}}
\def\vinj{v_r({\rm inj})}
\def\vzinj{v_z({\rm inj})}
\def\vrad{v_{\rm rad}({\rm inj})}
\def\hinj{H_{\rm inj}}
\def\csinj{c_s({\rm inj})}
\def\omegk{\Omega_{\rm K}}
\def\lk{l_{\rm K}}
\def\rsh{r_{\rm sh}}
\def\alf{\alpha}
\def\alfcr{\alpha_{\rm cr}}

\def\lsim{\lower.5ex\hbox{$\; \buildrel < \over \sim \;$}}
\def\gsim{\lower.5ex\hbox{$\; \buildrel > \over \sim \;$}}
\def\simeq{\lower.3ex\hbox{$\; \buildrel \sim \over - \;$}}

\shorttitle{Quasi-Spherical, Time-Dependent Viscous Accretion Flow}
\shortauthors{Lee et al.}

\begin{document}

\title{SIMULATIONS OF VISCOUS ACCRETION FLOW AROUND BLACK HOLES IN TWO-DIMENSIONAL CYLINDRICAL GEOMETRY}

\author{Seong-Jae Lee$^{1}$, Indranil Chattopadhyay$^2$, Rajiv Kumar$^2$, Siek Hyung$^1$, and Dongsu Ryu$^3$}
\affil{$^1$ School of Science Education, Chungbuk National
University, Chungbuk 28644, Korea seong@chungbuk.ac.kr\\
$^2$ ARIES, Manora Peak, Nainital-263002, Uttarakhand, India \\
$^3$Department of Physics,
School of Natural Sciences
UNIST, Ulsan 44919, Korea\\}

\begin{abstract}
We simulate shock-free and shocked viscous accretion flow onto a black hole in a
two dimensional cylindrical geometry, where initial conditions were chosen from analytical solutions.
The simulation code used the Lagrangian Total Variation Diminishing (LTVD) and remap routine, which enabled us
to attain high accuracy in capturing shocks and to handle the angular momentum distribution correctly.
Inviscid shock-free accretion disk solution produced a thick disk structure,
while the viscous shock-free solution attained a Bondi-like structure, but in either case, no jet activity nor any
QPO-like activity developed.
The steady state shocked solution in the inviscid, as well as, in the viscous regime, matched theoretical predictions well.
However, increasing viscosity renders the accretion shock unstable. Large amplitude shock oscillation is accompanied by
intermittent, transient inner multiple shocks. Such oscillation of the inner part of disk is interpreted as the source of
QPO in hard X-rays observed in micro-quasars.
Strong shock oscillation induces strong episodic jet emission. The jets also showed existence of shocks, which are produced
as one shell hits the preceding one.
The periodicity of jets and shock oscillation were similar.
The jets for higher viscosity parameter are evidently stronger and faster.
\end{abstract}


\section{INTRODUCTION}

Investigation of the behavior of matter and radiation around black holes gained popularity
when the accretion activity onto black holes became
the only viable model to explain the power and spectra radiated by various Active Galactic Nuclei
(AGN) and micro-quasars. Spectra
around black hole candidates (BHCs) in both AGNs and micro-quasars
show a thermal multi-colored component and non-thermal components. Some of these
BHCs show only non-thermal spectra
which can be fitted with the combination of one or two spectral indices, while
others require a combination of thermal and
non-thermal components. Moreover, most of these objects tend to be associated with
relativistic jets, and observations
indicate that these jets originate from within few tens of Schwarzschild radii
\citep{jbl99}.
Quasi-steady, mildly relativistic jets have been
observed in the hard spectral state
of the BHCs \citep{gfp03}, however, the jet power increases in the transient
outbursting objects, as they move from hard spectral states to intermediate states \citep{fbg04}.
Interestingly, the light curves of the stellar mass BHCs often show quasi
periodic oscillations (QPOs) of the hard photons
\citep{mkint92,mrg97,rsmm02,rmmo02,rm06,ndmc12}. Moreover, it has been shown that the evolution of
spectral states, QPOs and jet states
can be expressed by a hysteresis kind of a hardness-intensity diagram (HID), or Q diagram, and many of the
micro-quasars seem to follow the general pattern \citep{fbg04}.  It is to be noted, that any model
invoked to describe accretion-ejection mechanism around BHCs, should incorporate all of these issues.

Since the inner boundary condition for a black hole accretion has to be
supersonic, the first model of accretion onto a black hole was that of spherical radial inflow or relativistic Bondi
accretion \citep{m72}. However,
it was almost immediately pointed out that spherical accretion is too fast,
and therefore that matter does not have enough
time to produce the high luminosities outside the BHCs that are observed \citep{s73a, s73b}.
The focus then shifted to rotation dominated disk models which are optically thick
but geometrically thin
and with negligible radial infall velocity.
This disk model is called the
Shakura-Sunyaev disk, or, the
Keplerian disk \citep{ss73, nt73}. In spite of its simplicity, the Keplerian disk
model explained the `big blue bump',
or, the modified black-body part of the spectra from AGNs. However,
there are some theoretical deficiencies in purely Keplerian disks, because the inner boundary of a Keplerian disk
is too arbitrary, while the pressure gradient term is poorly treated. In addition, observationally the
Keplerian disk cannot explain the
presence of the hard power law tail. It was understood that a hot component closer to the horizon
could, in principle, scatter up the softer photons through an inverse-Compton process which would
explain the observed hard power law tail \citep{st80}. Since matter with non-negligible
advection is also hotter, various models emerged, which have a significant advection term along with rotation.

\citet{lt80} showed that an inviscid, rotating accretion flow, which is a
simpler form of advective flow, will have more than one sonic point.
Such accretion flows with multiple sonic points may undergo shock
transition both in inviscid, as well as, in the viscous regime \citep{f87,c89,c96}.

Aside from fixed $\gamma$
equation of state of the flow,
shocks have been obtained for flows with variable
$\gamma$ equation of state as well \citep{f87,cc11,kscc13,kc14}.
In the Paczy\'nski-Wiita pseudo potential domain \citep{pw80}, accretion shocks were
reported for various types of viscosity
prescriptions, like Chakrabarti-Molteni type \citep{c96}, Shakura-Sunyaev type
\citep{bdl08,kc13} and even for causal viscosity type \citep{gl02}. Accretion shocks
were reported for general-relativistic viscous disk as well \citep{ck16}.

However, the most popular of all the models in the advective regime is called
the advection dominated accretion flow (ADAF), which is characterized by
a single sonic point close to the horizon, and is subsonic further out
\citep{nkh97}.
ADAF, which was originally thought to be entirely subsonic and self-similar,
was found to be self-similar only at large
distances away from the horizon \citep{cal97}. More interestingly, the ADAF has been proved to be
a subset of a general advective solution \citep{lgy99,bdl08,kc13,kc14}.

Since the entropy of the post-shock flow is higher, the accretion flow would undergo shock
transition whenever such a possibility arise, because nature favors higher entropy solution.
Shocks in accretion disks around black holes are advantageous.
The post-shock disk (PSD) is
hotter, slower, and denser than the pre-shock flow, although the density is not high enough to make the PSD optically
thick. Hence, the PSD
acts as a hot Comptonizing
cloud that would produce the inverse-Comptonized hard power law tail. The Comptonizing
cloud obtained in this manner is not an arbitrary addition on the top of a disk solution, but
comes naturally by solving the
equations of motion in the advective regime, as would be shown in this paper as well.
In a model solution, \citet{ct95} solved the radiative transfer
equation for a two component accretion flow, involving matter with high viscosity and Keplerian
angular momentum distribution, as well as, sub-Keplerian matter. Matter with local Keplerian angular momentum
occupies the equatorial plane and the sub-Keplerian flow sandwiches the Keplerian disk from the top and bottom.
The sub-Keplerian flow, being hot and supersonic, experiences a shock transition, and as a result supplies hot
electrons. The Keplerian disk supplies soft photons. The post-shock flow, being hot and puffed up
intercepts soft photons from the Keplerian disk, inverse-Comptonize them to produce the hard power-law
tail as is observed in the low-hard spectral state of the micro-quasars \citep{ct95,mc10,gc13}.
If the Keplerian accretion rate is increased beyond a critical limit, it cools down the
post-shock disk, creating what is known as the high-soft spectral state. Recent simulations
show that this scenario is a distinct possibility \citep{gc13}.

Interestingly, the PSD may or may not be stationary. The
PSD may be subject to a large number of instabilities.
Since the PSD is hotter and denser, the cooling time scales may or may not
be comparable with the dynamical time scale; and where the two are comparable,
the shock may oscillate \citep{msc96b,otm07}.
And since the PSD produces the high energy power-law tail of the radiation spectrum,
the oscillating shock should induce the same oscillation in hard photons too
--- a very natural explanation of QPO in micro-quasars. The persistent oscillation, or,
instability of the PSD is not only related to the resonance between cooling and infall time scales,
but viscosity might induce shock oscillations as well \citep{lmc98,lrc11,dcnm14}.
There have been many stability studies of shock \citep{n92,n94,nh94,gf03,gl06},
but it was shown that even under non-axisymmetric perturbations, the shock tends to persist,
albeit, as a deformed shock \citep{mtk99}.

Apart from explaining the origin of hard power-law radiations and origin of the QPO,
the extra thermal gradient force in the PSD powers bipolar outflows.
These outflows may be considered as precursor of jets \citep{mlc94,mrc96a,msc96b,cd07,otm07,kc13}.
The HID for micro-quasars shows that as the micro-quasar enters the outbursting stage,
both QPO and jet power increase while spectral state evolves from low hard to intermediate hard/soft state \citep{fbg04,rn13}.
Interestingly, since the post-shock region of the disk generates the outflow and also shocks form close to the black hole,
the observational constraint that a jet base is formed close to the horizon \citep{jbl99} is also satisfied.
Recently, \citet{kcm14} showed that if the radiative acceleration of the shock-driven outflows are considered, then
jet power increases as the spectral state of the disk moves from low hard to intermediate hard states,
exactly confirming the fact that has been observed \citep{fbg04}.

Numerical simulations of accretion disks around black holes, have been performed
with codes based on smooth particle hydrodynamics (SPH),
which has higher artificial viscosity \citep{mlc94,dcnm14}, whereas others
with Eulerian codes \citep{mrc96a,ny09,otm07}. Eulerian codes are based on upwind schemes
and conserve energy and momentum naturally. So they efficiently capture and solve the discontinuities like shock waves.
However, in Eulerian schemes, azimuthal momentum is conserved but angular momentum component is not.
SPH code, on the other hand, conserves angular momentum accurately in absence of viscosity.
\citet{lrc11} developed a TVD plus remap method, which combines the Lagrangian method and TVD method
efficiently. With this Lagrangian TVD (LTVD) code, shocked accretion and ADAF type solutions were
accurately reproduced, and the code strictly conserves angular momentum in the inviscid scenario.
Using one dimensional LTVD code, \citet{lrc11} accurately reproduced theoretical accretion solutions,
with strict conservation
of angular momentum in inviscid flow. Introduction of viscosity creates a situation that
the angular momentum redistributes and its dissipation becomes accentuated.
As a result, beyond a critical value of viscosity the PSD starts to oscillate.
Moreover, the possibility of forming multiple shocks, or, shock cascade conjectured by \citet{ft04}
were also obtained in \citet{lrc11}, and shocks were observed to oscillate with separate, distinct
frequencies.

In one-dimensional simulations, the dynamics in the vertical direction is suppressed. Therefore, the
accretion-ejection phenomena cannot be investigated, because the ejection occurs in the vertical direction away
from the equatorial plane. In this paper, we
follow the methods of \citet{lrc11} and perform multi-dimensional simulations of viscous accretion flow.
Although shocks form for an inviscid accretion flow, is it possible to find steady shocks for a high
viscosity parameter?
Do multiple shocks form for multi-dimensional simulations, or is the phenomenon an artifact
of one dimension? 
Moreover, earlier multi-dimensional simulations show that the shock leaves the computational domain
for higher viscosity parameter \citep{lmc98}. The consensus reached was that, for higher viscosity in the flow,
shock withers away. In an one dimensional simulation of \citet{lrc11}, the shock went out of the simulation box
for high viscosity. However, in the one dimensional analysis dynamics along other directions are suppressed,
therefore, exaggerated dynamics along the relevant direction may force the shock to leave the computational domain.
In this paper, we would like to study the fate of the shock
in multi-dimensional simulations for higher viscosity.
In order to accommodate for large amplitude shock oscillations, we have chosen a larger
computational box. Moreover, do the bipolar outflows from the PSD leave the computational
domain with significant velocities in order to qualify these outflows as jet precursor?
In section 2, we present the governing equations. In section 3, we describe the code
and the tests performed to check the veracity of the code in multi-dimensions. In section
4, we discuss the theoretical results and comparing with simulations. In  section 5, we discuss the temporal
behavior of a viscous accretion disk. In the last section, we present concluding remarks.

\section{BASIC EQUATIONS}

The mass, momentum, and
energy conservation equations in two-dimensional cylindrical coordinates ($r$, $\theta$, $z$) are given by

\begin{equation}
\frac{\partial \rho}{\partial t} +\frac{1}{r}\frac{\partial (r\rho v_r)}{\partial r}
+ \frac{\partial (\rho v_z)}{\partial z} = 0,
\label{cont.eq}
\end{equation}
\begin{equation}
\frac{\partial (\rho v_r)}{\partial t} + \frac{1}{r} \frac{\partial(r \rho v_r^2)}{\partial r}
+ \frac{\partial(\rho v_r v_z)}{\partial z} + \frac{\partial P}{\partial r} = -\rho \frac{\partial\Phi}{\partial r}
+ \frac{\rho l^2}{r^3},
\label{mr.eq}
\end{equation}

$$
 \frac{\partial (\rho v_{\theta})}{\partial t} + \frac{1}{r} \frac{\partial(r \rho v_{\theta} v_r)}{\partial r}
+ \frac{\partial (\rho v_{\theta} v_z)}{\partial z}  
$$
\begin{equation}
= \frac{1}{r^2} \frac{\partial}{\partial r}
\left(\mu r^3\frac{\partial \Omega}{\partial r}\right) + r\frac{\partial}{\partial z}
\left(\mu \frac{\partial \Omega}{\partial z}\right),
\label{mfi.eq}
\end{equation}
\begin{equation}
\frac{\partial (\rho v_z)}{\partial t} + \frac{1}{r} \frac{\partial (r\rho v_r v_z)}{\partial r}
+ \frac{\partial (\rho v_z^2)}{\partial z} + \frac{\partial P}{\partial z} = - \rho \frac{\partial \Phi}{\partial z},
\label{mz.eq}
\end{equation}
$$
\frac{\partial {\bar E}}{\partial t} + \frac{1}{r} \frac{\partial}{\partial r}(r{\bar E}v_r) +\frac{\partial}{\partial z}
({\bar E}v_z) +
\frac{1}{r}\frac{\partial}{\partial r}(rPv_r) + \frac{\partial}{\partial z}(Pv_z)
$$
\begin{equation}
= \frac{1}{r} \frac{\partial}{\partial r}\left(r^2 \mu v_\theta \frac{\partial \Omega}{\partial r} 
\right) + \frac{\partial}{\partial z}\left(\mu v_\theta \frac{\partial v_\theta}{\partial z}\right)
-\rho v_r \frac{\partial \Phi}{\partial r} - \rho v_z \frac{\partial \Phi}{\partial z},
\label{energ.eq}
\end{equation}
where, $\rho$, $v_r$, $v_{\theta}$, $v_z$, $l$, $\Phi$ and ${\bar E}$ are the gas density, radial velocity,
azimuthal velocity, vertical velocity, specific angular momentum, gravitational potential, and total energy
density, respectively. Here, ${\bar E}=\rho (v^2_r+v^2_{\theta}+v^2_z)^{1/2}+\rho e$. Axis-symmetry is assumed.
The angular velocity is defined as $\Omega = v_{\theta}/r=l/r^2 $ and the pseudo-Newtonian
gravity \citep{pw80} assumed to mimic the Schwarzschild geometry, is given by:
\begin{equation}
\Phi = -{G\mbh \over R - \rg};~~\mbox{where, } R=\sqrt{r^2+z^2}
\end{equation}
where $\mbh$ is the black hole mass and the Schwarzschild radius is
$\rg=2G\mbh/c^2$.
The gas pressure in the equation of state for ideal gas is assumed,
\begin{equation}
P = (\gamma-1) \rho e,
\end{equation}
where $\gamma$ is the ratio of specific heats.
Shakura \& Sunyaev's viscosity prescription ($\alpha$) is assumed,
\ie\ the dynamical
viscosity coefficient is described by
\begin{equation}
\mu = \alpha \rho {c_s^2 \over \omegk},
\label{viscdyn.eq}
\end{equation}
where, the viscosity parameter $\alpha \leq 1$ is a constant.
The square of the adiabatic sound speed is given by,
\begin{equation}
c^2_s=\frac{\gamma P}{\rho}
\end{equation}
and
\begin{equation}
\omegk = {\lk \over r^2}= \left[\frac{1}{r}\frac{\partial \Phi}
{\partial r}\right]^{1/2}
\end{equation}
is the Keplerian angular velocity.
We have ignored cooling in this paper. We have assumed that only the $r-\theta$ component of the viscous stress tensor is
significant.

In the following, $\mbh$, $c$ and $\rg$ are used as the units of mass, velocity and length,
respectively. Therefore, the unit of time is $\tg=\rg/c$. All of the equations, then, become dimensionless by using the above unit system.

\section{CODE}

One of the most demanding tasks in carrying out numerical
simulations of transonic flow is to capture shocks sharply.
The upwind finite-difference schemes on an Eulerian grid have been known to
achieve the shock capture strictly. However, since the angular momentum of
equations (\ref{cont.eq})--(\ref{energ.eq})  is not treated as a conserved quantity in such Eulerian
codes, we use the so-called LTVD scheme. The newly designed
code can preserve the angular momentum perfectly because the Lagrangian concept
is used, and it can also guarantee the sharp reproduction of discontinuities
because the TVD scheme \citep{h83,rokc93} is also applied
\citep[see][for details]{lrc11}.
The calculation in the
angular momentum transfer is updated through an implicit method, assuring
it is free from numerical instabilities related to it. But the
viscous heating without cooling is updated with a second order explicit method,
since it is subject to less numerical instabilities.

\subsection{Hydrodynamic Part in Multi-Dimensional Geometry}

We start with the hydrodynamic part in Lagrangian step and remap,
which consists of plane parallel and cylindrical geometry.
The conservative form of equations (\ref{cont.eq})--(\ref{energ.eq}), in mass coordinates and in the
Lagrangian grid, can be written as:

\begin{equation}
{d\tau \over dt} - {\partial(r^{\tilde{\alpha}} v) \over \partial m} = 0,
\label{lcont.eq}
\end{equation}
\begin{equation}
{dv \over dt} + r^{\tilde{\alpha}}{\partial p \over \partial m} = 0,
\label{lmr.eq}
\end{equation}
\begin{equation}
{dl \over dt} = 0,
\label{lmfi.eq}
\end{equation}
\begin{equation}
{dE \over dt} + {\partial(r^{\tilde{\alpha}} vp) \over \partial m} = 0,
\label{lenerg.eq}
\end{equation}
where ${\tau}$ and $E$ are the specific volume and the specific total
energy, respectively, related to the quantities used in equations
(\ref{cont.eq})--(\ref{energ.eq}) as
\begin{equation}
\tau = {1 \over \rho}, \qquad E = e + {v^2 \over 2}.
\label{mascord.eq}
\end{equation}
The mass coordinate related to the spatial coordinate is
\begin{equation}
dm = \rho(r) r^{\tilde{\alpha}} dr,
\label{mascord2.eq}
\end{equation}
and its position can be followed with
\begin{equation}
{dr(z) \over dt} = v(m,t)
\label{mascord3.eq}
\end{equation}
where ${\tilde{\alpha}}$ represents the parameters in different geometrical
geometry; \ie ${\tilde{\alpha}} = 0$ refers to the cartesian coordinate system, while
${\tilde{\alpha}} = 1$ refers to cylindrical geometry. Since the equations (11), (12),
and (14) show a hyperbolic system of conservation equations, upwind schemes are
applied to build codes that advance the Lagrangian step using the Harten's
TVD scheme \citep{h83}. Since the conserved equations (\ref{cont.eq})--(\ref{energ.eq}) are decomposed into one-dimensional functioning code through a Strang-type directional splitting \citep{st88} like in \citet{ryu95}, ${\tilde{\alpha}} = $ 1 \&  $v=v_r$ and  ${\tilde{\alpha}} = $ 0 \& $v=v_z$ are used for calculations along the r and z directions, respectively. And $v_{\theta}$ is handled separately.
The detailed explanations of Lagrangian TVD and remap are given in \citet{lrc11}.
The equation (\ref{lmfi.eq}) does not need to be updated in the Lagrangian
step since it is preserved in the absence of viscosity. Equations (\ref{cont.eq})--(\ref{energ.eq}), calculated by the Lagrangian and remap steps,
are updated in the Eulerian grid except for the centrifugal
force, gravity, and viscosity terms on the right-hand side. The
centrifugal force in the r direction only and gravity terms in the r and z directions are calculated separately
after the Lagrangian and remap steps such that

\begin{equation}
v_i^{\rm hydro} = v_i^{\rm lag+remap} + \Delta t \left( {l_i^{\rm remap}
\over r_i^3} - {d\Phi \over dr(z)}\Bigg|_i \right).
\end{equation}

\noindent Then, the viscosity terms are calculated, as discussed in the
following subsection.

\subsection{Viscosity Part}

Viscosity plays an important role in transferring  angular momentum
outwards and it allows the matter to accrete inwards around a black hole.
The angular momentum transfer in equation (\ref{mfi.eq}) is described by the viscosity
parameter given in \citet{ss73}.

Since the terms for the angular momentum transfer of radial and vertical
directions in equation (\ref{mfi.eq}) are linear
in $l$, it is possible to calculate implicitly.
Substituting $(l^{\rm new} + l^{\rm remap})/2$ for $l$, equation (\ref{mfi.eq}) without
the
advection term becomes
$$
a'_i l_{i-1}^{\rm new} + b'_i l_{i}^{\rm new} +c'_i l_{i+1}^{\rm new} =
$$
\begin{equation}
-a'_i l_{i-1}^{\rm remap} - (b'_i -2)l_{i}^{\rm remap} -c'_i l_{i+1}^{\rm remap},
\end{equation}
forming a tridiagonal matrix. Here $a'_i$, $b'_i$, and $c'_i$ are given
with $\rho$, $\mu$, and $r$ as well as $\Delta r$, and $\Delta t$, while $a_i$, $b_i$, and $c_i$ are given
with $\rho$, $\mu$, and $z$ as well as $\Delta z$, and $\Delta t$.
The tridiagonal matrix is solved properly for $l^{\rm new}$ with
appropriate boundary conditions \citep{ptvf92}.
Another role of viscosity is to act as friction resulting in viscous heating.
Here, the viscous heating energy is
fully saved as an entropy, since we ignore cooling. Our experience in
dealing with numerical experiments tells
us that the explicit treatment for the calculation of the viscous heating term does
not cause any numerical problems. Thus
angular momentum transfer is solved implicitly, while frictional heating energy
is solved explicitly.

\section{FORMATION OF SHOCK IN TWO DIMENSIONAL GEOMETRY}

\subsection{To Regenerate Two Dimensional Simulation Solution - A Test for the Code}

\begin{figure}[p]



\plotone{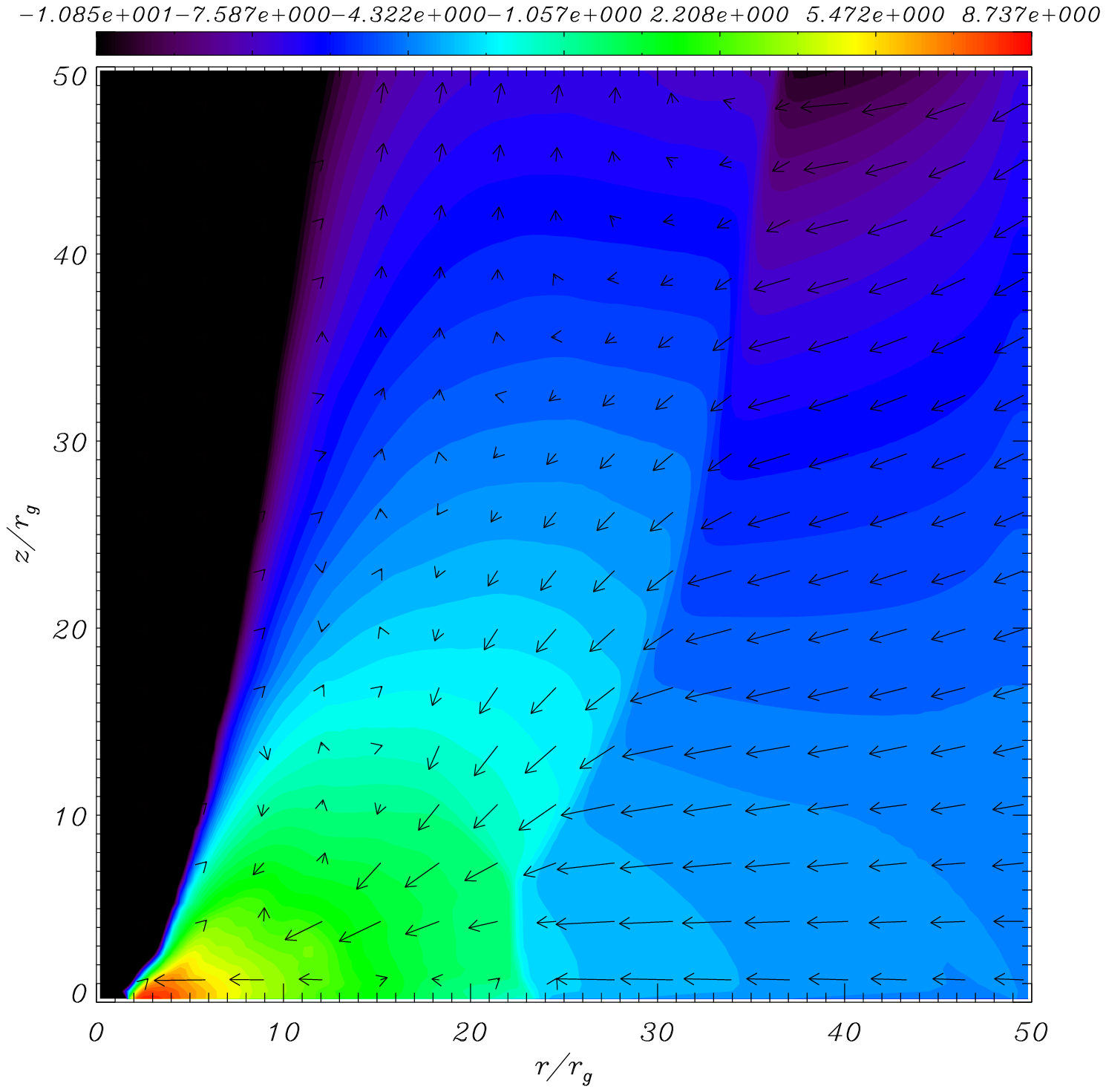}
\caption{Density contours and velocity fields of a shocked accretion flow in the
r-z plane.
The big paraboloidal accretion shock touches the equatorial plane at $r \sim
24\rg$. The flow parameters
are $v_{\rm rad} = -0.068212c$, $c_s = 0.061463c$, $\gamma=4/3$, and $l=1.65\rg c$.}
\label{lab:fig1}
\end{figure}

We present one of test results to demonstrate that the code can capture shocks sharply
and resolve the structure clearly in a
transonic flow. In the test, our result, in fact, corresponds to one of the earlier simulation results by \citet{mrc96a}.
The inviscid flow with the same initial conditions as in \citet{mrc96a} enters
from the outer boundary,
e.g., $\vrad \left[\equiv {\sqrt{\vinj^2 + \vzinj^2}}\right] = -0.068212c$, sound speed
$\csinj = 0.061463c$, adiabatic index $\gamma = 4/3$,
and specific angular momentum $l = 1.65\rg c$. The calculation in cylindrical
geometry was performed with 128 x 256
cells in a 50 x 100$\rg$ box size. Figure \ref{lab:fig1} clearly shows the
presence of one shock structure along the
equatorial plane, as seen in the result calculated by the smoothed particle
hydrodynamic (SPH) technique. Here the shock is
resolved sharply as seen in the result calculated by the Eulerian total variation
diminishing (TVD) technique. Since the present code
uses the Lagrangian scheme, in absence of viscosity it can conserve angular momentum
strictly. Hence, we can minimize the errors of
the calculation of the specific angular momentum, present in a purely
Eulerian scheme.

\subsection{Theoretical Steady State Solutions}

So far, obtaining a proper time dependent accretion solution around black holes
is possible only through numerical simulations.
However, early notions of the accretion-ejection paradigm
emerged through theoretical efforts for semi-analytical solutions of the governing equations
(\ref{cont.eq})--(\ref{energ.eq}),
in steady state (the so-called 1.5-dimensional analysis).
In steady state, the governing equations (\ref{cont.eq})--(\ref{energ.eq}) for the disk can be
integrated to obtain the following
constants of motion \citep{kc13}, where the mass accretion equation is
\begin{equation}
 {\dot M}=4 \pi r H \rho v_r,
\label{accrt.eq}
\end{equation}
and the specific energy, or the generalized Bernoulli parameter for viscous flow is
\begin{equation}
 {\cal E}_{\rm g}=0.5v^2_r+\frac{c^2_s}{(\gamma-1)}-\frac{l^2}{2r^2}+\frac{ll_0}{r^2}+\Phi.
\label{bern.eq}
\end{equation}
Here $l_0$ is the specific angular momentum on the horizon---an integration constant,
and $H={\sqrt{(2/\gamma)}}c_sr^{1/2}(r-1)$
is the local half height of accretion disk, assumed to be in hydrostatic equilibrium along the vertical direction.
The gradient of the angular velocity obtained by integrating the azimuthal component of Navier Stokes
equation as per the assumptions is given by
\begin{equation}
 \frac{d\Omega}{dr}=-\frac{\gamma v_r \Omega_K(l-l_0)}{\alpha c^2_s r^2}.
 \label{domegadr.eq}
\end{equation}
It is very clear that in the absence of viscosity ($\alpha=0$), $l=l_0$, and therefore, equation (\ref{bern.eq})
takes the usual form of Bernoulli parameter ${\cal E}={\cal E}_{\rm g}=0.5v^2_r+c^2_s/(\gamma-1)+l^2_0/(2r^2)+\Phi$.
Now for a given value of ${\cal E}_{\rm g}$, $l_0$, and $\alpha$, the entire steady state solution in
1.5-dimension is obtained. In the rest of the paper we have assumed $\gamma=1.4$, a value which will approximately
describe electron-proton flow close to the horizon \citep{cr09,cc11,kscc13,kc14}. In this paper, we have used inviscid
analytical solutions as initial conditions for viscous flow.

\subsection{Comparison of Numerical Simulation with Theoretical Inviscid Solutions}

Next we compare solutions obtained from our simulation code with analytical results of \citet{kc13}.
We compare shock-free, as well as, shocked accretion solution.
The accreting flow is supplied from the outer
boundary which will be mostly absorbed at the inner edge of an accretion disk. The behavior of inviscid
accreting matter around a black hole depends on the initial parameters of
inflow, for instance, its specific energy ${\cal E}$ and specific angular momentum $l_0$
\citep{kc13}.
As mentioned before, the theoretical steady state solutions are obtained for a 1.5-dimensional analysis
\ie a disk assumed to be in vertical hydrostatic equilibrium, while the simulation
is done properly in two spatial dimensions. For $\gamma=1.4$ and 1.5-dimension, steady state
shock solution exists for $1.5\rg c<l_0<1.8\rg c$ \citep[for details, see Figure 2 of][]{kc13}.
We choose two analytical
solutions from \citet{kc13}: model one, or M1, is a theoretical ``shock-free'' accretion solution
with parameters, $l_0 = 1.48\rg c$, and the specific energy ${\cal E}=0.0063c^2$.
The inflow variables at the injection radius
$\rinj=200\rg$: $\linj=1.48\rg c$, $\vinj=-6.955509\times 10^{-3}c$, $\vzinj=0$,
and
$\csinj=5.920845\times10^{-2}c$.
The computational box size is $200\rg~\times~200\rg$
with the resolution of $400\times400$ cells.

\begin{figure}
\plotone{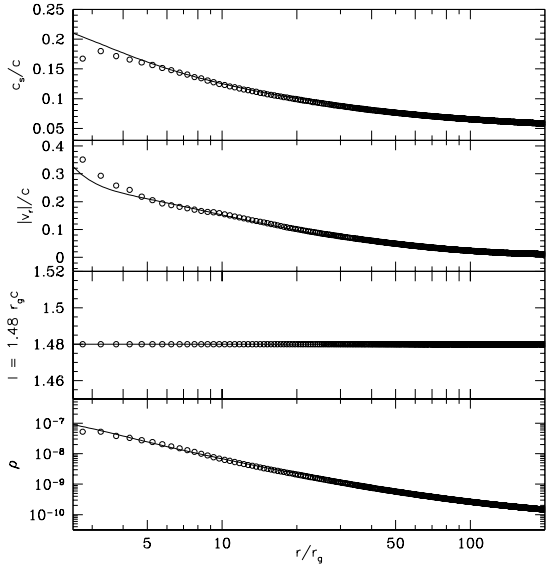}
\caption{Test of the shock-free solution of case M1: $\rinj=200$, $\vinj=-6.955509\times 10^{-3}c$,
$\csinj=5.9200845\times 10^{-2}c$ and $\linj = 1.48$
$\rg c$.
The solid lines represent the analytical solution, while the open circles represent
the numerical solution.
The adiabatic sound speed $c_s$, radial velocity $v_r$, specific angular
momentum $l$, and density
$\rho$ along the equatorial plane are shown from top to bottom.}
\label{lab:fig2}
\end{figure}

Figure \ref{lab:fig2} compares simulation (open circles)
with analytical (solid line) solutions, which represent sound speed, radial velocity,
specific angular momentum, and density distribution along the equatorial plane from top to
bottom. The simulation rigorously regenerates the analytical no-shock solution, once steady state is reached.
The agreement between the simulation and the analytical solution is remarkable.
Close to the horizon, the flow falls very fast onto the black hole,
so the vertical equilibrium assumption is not strictly maintained in those regions, causing
a slight mismatch of $v_r$ and $c_s$ with
the theoretical solution.

\begin{figure}

\plotone{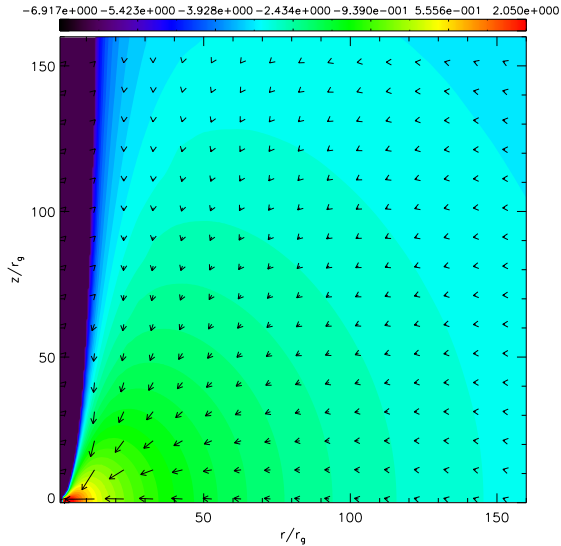}
\caption{Density contour map and velocity field of the shock-free
case M1.}
\label{lab:fig3}
\end{figure}

Figure \ref{lab:fig3} shows the density contour (color gradient) and velocity field (arrows) from the simulation
of case M1 in the $r-z$ plane. Interestingly, the density contours mimic the thick
disc \citep{pw80} configuration, although the advection term is significant in this simulation.
We then simulate with injection parameters taken from \citet{kc13}, which predicts
theoretical shock in the inviscid limit, and we call this case M2. The parameters of M2 correspond
to ${\cal E}=1.25\times10^{-5}c^2$ and $\linj=1.7\rg c$, with injection parameters
$\vrad=-4.249299\times 10^{-2}~c$, $\csinj=1.190908\times 10^{-2}~c$ at $\rinj=400\rg$.
The height of the disc at $\rinj$ is $H_{\rm inj}=113.75\rg$.

\begin{figure}


\plotone{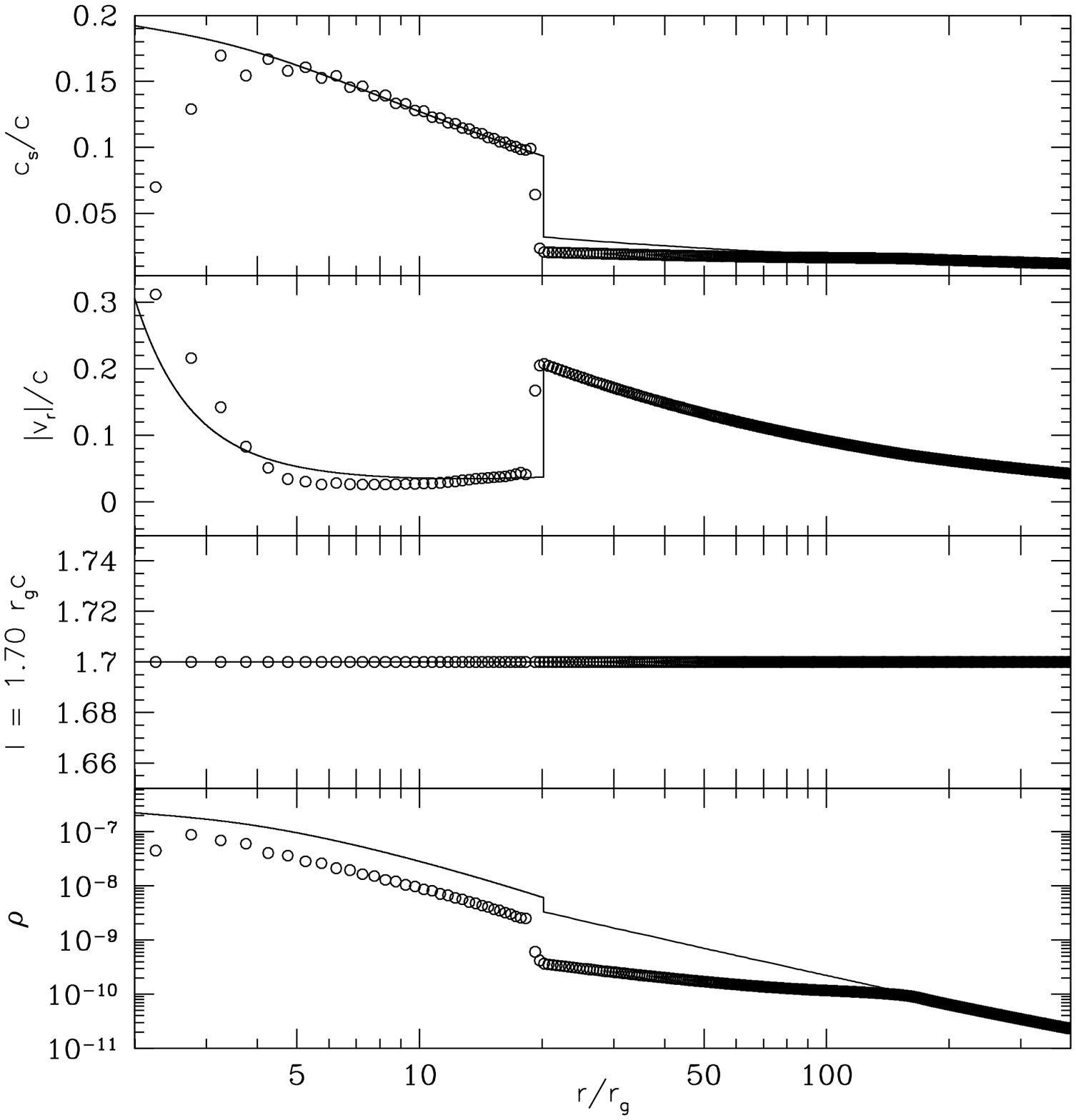}
\caption{
Case M2: The injection parameters here are the injection
radius $\rinj=400\rg$, $\vinj
=-4.249299 \times 10^{-2}c$, $\csinj= 1.190908\times 10^{-2}c$, $\linj=1.7 \rg c$ and
the height at $\rinj$ is $H_{\rm inj}=113.75~\rg$.
The sound speed,
velocity, specific angular momentum, and density are shown from top to bottom.
The solid lines and open circles represent the analytical solutions and the
numerical results, respectively. The analytical shock location is at 20.18
$\rg c$, while the numerical one is at 19.25 $\rg c$.}
\label{lab:fig4}
\end{figure}

Figure \ref{lab:fig4} shows the sound speed, radial velocity, specific angular momentum,
and density distribution along the equatorial plane which are plotted in panels from top to bottom, respectively. The solid lines show
the analytical solution while the open circles show numerical results for the M2 solutions. 
The computational box size is $400 \times 200 ~\rg$ with $800 \times 400$ cells. 
The shock location from numerical calculations along the
equatorial plane is about $19.25\rg$, while the shock position suggested by
the analytical solution is $20.18\rg$.
The agreement of theoretical solution (solid) with the numerical one (hollow circles) is fairly
remarkable, for the simple reason the numerical result is not restricted to no out-flow and
vertical hydrostatic equilibrium, while the theoretical result is.
Since hydrostatic equilibrium is, however, not well maintained close to the horizon,
the shock location in the equatorial plane is slightly
closer to the horizon than the theoretically predicted value indicates.

\begin{figure}


\plotone{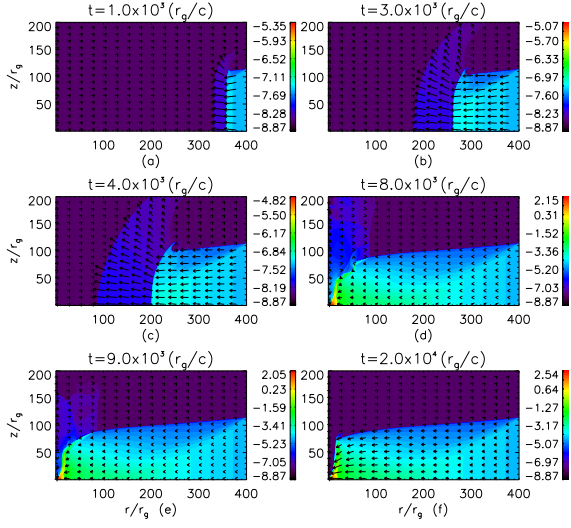}

\caption{Density contours and velocity fields
of a shocked inviscid disk. Six time steps (in units of dynamical time $\tg=\rg/c$)
are plotted to show how steady state is reached.
The initial conditions are the same as in Figure 4.}
\label{lab:fig5}
\end{figure}

Figure \ref{lab:fig5} shows the snapshots of density contour and velocity field of case M2 at six
time steps, showing how the solution progresses into steady state. The first snapshot
is for the time ($t=10^3 \tg$) when the
accreting matter is still far away from the horizon. In the second and the third
panels ($t=3\times10^3\tg$ and $t=4\times10^3\tg$), the injected matter
has still not reached the horizon. The fourth ($t=8\times10^4\tg$) and fifth ($t=9\times10^3 \tg$)
panels  show the formation of unsteady shocks with weak time-dependent
post-shock outflows. The shock becomes steady at $t>1.2\times10^4\tg$ as the solution reaches steady state.
Here, the density contours and velocity vectors are plotted for time $t=2\times10^4\tg$.
The inflow matter hits the effective potential
barrier and is piled up behind the barrier, where the accretion shock is formed.
The earlier theoretical work already showed that there are two shock locations \citep{f87,c89} where
the inner shock was found to be unstable while the outer one is stable \citep{n92,mlc94}.
In our study, we also observe that the shock actually forms closer to the horizon, but settles
around the stable outer shock location once the steady state is reached. In the rest of the paper,
we use the steady state solution of M1 \& M2 as the initial condition for the viscous flow.

\section{SIMULATION OF VISCOUS FLOW}

\subsection{Steady State shock-free disk}
\begin{figure}


\plotone{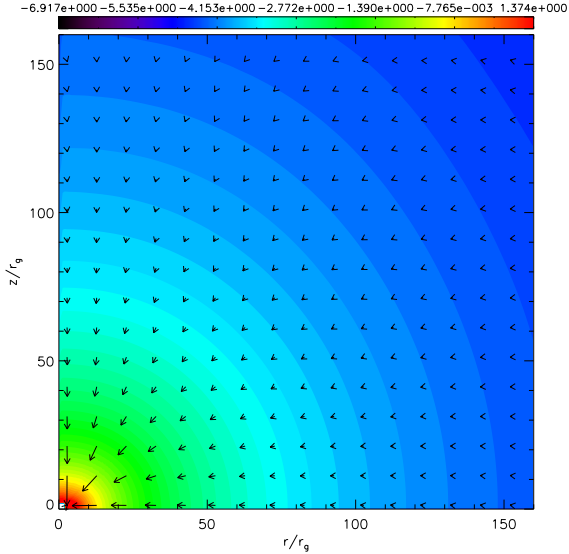}
\caption{Density contours and velocity fields
of a shock-free viscous disk for $\alpha=0.05$.
The initial conditions are the same as in Figure \ref{lab:fig2}, or, M1.}
\label{lab:fig6}
\end{figure}

We turn on viscosity on the steady state of M1, or, Figure \ref{lab:fig2}. Viscosity transports angular momentum,
and close to the horizon, the angular momentum decreases a lot and the disk morphology which represented that
of thick disk in the inviscid limit resembles more like a Bondi flow. The flow direction is essentially spherical radial, as is
seen from the velocity vectors of Figure \ref{lab:fig6} once steady state is reached, the density contours are almost spherical, corroborating
radial type or Bondi type flow. The viscosity in this case is $\alpha=0.05$, but we have also checked for $\alpha=0.1$
and it remains a Bondi type flow. No jet like structure is seen, and no instability is seen which can be treated as a source of
QPOs.

\subsection{Steady State shocked viscous disk}

In the next step, we include the viscosity terms to the aforementioned steady state solution of M2. With small
$\alpha$, the viscous solution remains stable, albeit for a different value of shock location, or, $\rsh$.
With the same injection parameters
as that of inviscid shocked flow, \ie M2: $\vrad=-4.249299\times 10^{-2}~c$,
$\linj=1.7 \rg c$
and $\csinj= 1.190908\times 10^{-2}~c$ at $\rinj=400\rg$,
we turn on the viscosity of $\alpha=0.002$ at $t\sim 2.6\times10^4 \tg$.
A theoretical solution with these injection parameters at $\rinj=400\rg$, corresponds to
a specific energy of ${\cal E}_g=1.25\times10^{-5}~c^2$ and $l_0=1.69966 \rg c$.

Figure \ref{lab:fig7} shows the corresponding global theoretical solution (solid) and the
equatorial values of the simulation result (open circles). Top three panels show the distributions
of $c_s$ in (a), $|v_r|$ in (b) and $l$ in (c), respectively, while Figure \ref{lab:fig7} (d)
shows the evolution of the equatorial shock location $\rsh$ obtained
from the simulation as a function of time. In the simulations, the steady shock location is at $\rsh=22.25 \rg$, while
the theoretical shock is obtained at $22.45\rg$. The position of $\rsh$ moves out as viscosity is turned on.
For low $\alf$, the angular momentum transport between $\rinj$ and $\rsh$ is negligible, so in the pre-shock
disk $l$ is roughly constant. It must be remembered though, if the computational box was increased to $10^5 \rg$,
then the variation of angular momentum would have been discernible, as is exhibited by the theoretical solution.
Since the PSD is much hotter, the angular momentum transport is more efficient for the same value of $\alf$. This
causes the local angular momentum in the PSD to be greater than $\linj$. The extra centrifugal force therefore pushes out the shock
front outwards.
Figures \ref{lab:fig7} (a)--(c) show the robustness of both the simulation and the analytical solution.

The PSD may eject outflows and experience turbulences, therefore some disagreements are inevitable
between the analytical and simulation results. Moreover, since the vertical assumption do not hold well near the horizon,
so close to the horizon both $c_s$ and $v_r$ deviates
from the analytical value. The angular momentum distribution
of the simulation deviates from the analytical indication in the
post-shock disk region. However, the maximum fractional departure of the angular momentum distribution
of the simulation from the analytically obtained value is $\Delta l_{\rm sim}/l_{\rm analy} \lsim 0.016$.
Such a small degree of the deviation is within acceptable limit, considering that the $\rsh$ is reproduced quite accurately.
We have plotted the analytical solution up to $r=10^5 \rg$, in order to show that $\rinj$ is not the actual outer
boundary. Since the simulation for an eigenvalue solution like that of the
accretion disk in a huge box of $10^5 \rg$ length scale is inconceivable or very expensive, we simulate the inner region of the disk.
It is advisable that one should be careful in analyzing or addressing the
outer boundary condition when the simulation
box is only within the inner few hundred Schwarzschild radii.

Figure \ref{lab:fig8} displays snapshots of density contours and velocity vectors of the flow with the same initial and boundary
conditions as in Figure \ref{lab:fig7}, at various time steps (marked above the panels). These snapshots show that indeed the
solution reaches the steady state at $t \gsim 4 \times 10^4 \tg$.
For both the viscous and inviscid cases, the agreement between theoretical/semi-analytical solution and the
simulated solutions on the equatorial plane is fairly satisfactory given the fact that analytical solutions
are obtained under vertical equilibrium and no outflow assumptions, while the simulations are just time dependent
solutions of the fluid equations in two dimensions, where such assumptions are not implemented. As far as we know,
the comparison of a theoretical solution and a simulated solution for a steady state shock in the presence of viscosity
was not much done in earlier studies.

\subsection{Shock Oscillation in a Disk}

Shock oscillations have been observed in the presence of cooling \citep{msc96b,otm07},
for inviscid and adiabatic flows and for Newtonian point mass gravity \citep{rbol95}, or,
for stronger gravity \citep{rcm97}, also in presence of viscosity \citep{lmc98,lcscbz08,lrc11,dcnm14} etc.
It has been generally accepted that accretion shocks may exist for low viscosity and
cannot be sustained
for $\alpha >~{\rm few}\times 10^{-3}$ \citep{lmc98,lcscbz08}.
However, shocks may exist theoretically for $\alpha \lsim 0.3$ \citep{kc13,kc14}, which is fairly high.
The flow parameters we have chosen for our simulation, are in the domain where steady shocks do not exist
for high $\alpha$. We would therefore like to find out whether oscillatory shocks exist for these injection
parameters, or the shock completely fades away.
With an one dimensional LTVD code, we showed that persistent oscillatory shocks exist for $\alpha \sim$ few$\times 10^{-2}$
\citep{lrc11}. Presently we would like to investigate this scenario in multi-dimensions, since LTVD as a scheme is superior
to both TVD, as well as, Lagrangian code.

As has been mentioned, the initial condition for the viscous flow is the steady state
as in M2, and the boundary condition of M2 is also employed.
In our study, we found out that the steady state shock tends to oscillate
for $\alpha>0.003$. Our results also show that a hotter PSD ensures higher average $l$ than that of the immediate pre-shock disk.
This causes an outward centrifugal thrust which pushes $\rsh$ out. If this thrust is greater than the sum of ram pressure and the
gas pressure of the pre-shock disk then $\rsh$ will move out instead of settling down. However, the expanding $\rsh$ also causes a total pressure drop within PSD. This would restrict the outward motion trying to contract $\rsh$. Due to the competition between outward expansion and contraction, the $\rsh$ is in oscillation mode.
In Figure \ref{lab:fig9}, we plot $\rsh$ with $t$ for (a) $\alpha=0.003$, (b) $0.005$,
(c) $0.007$ and (d) $0.01$, respectively. The shock starts to oscillate as in Figure \ref{lab:fig9} (a), and then undergoes close to a regular oscillation for higher $\alpha$ (b). In the case of higher $\alpha$ (c) and (d), the shock oscillation is not in regular mode any more and the
amplitude of oscillation increases.

Figure \ref{lab:fig10} shows the snapshots of density contours and velocity field of an accretion solution
for $\alpha=0.01$. The time of each snapshot is mentioned in the figure. For $\alpha=0.01$ the jets
are observed to be episodic.
The strength of the jet is clearly related to the dynamics of PSD, but now multiple
shocks appear. In order to show multiple shocks, we plot $-v_{\rm r}/c$ (Figure \ref{lab:fig11}
a--d), $c_s/c$ (Figure \ref{lab:fig11}
e--h) and $l/(\rg c)$ (Figure \ref{lab:fig11} i--l), measured on the
equatorial plane, at $t=2.474\times10^5 \tg$ (Figure \ref{lab:fig11} a, e, i), $t=2.480\times10^5 \tg$
(Figure \ref{lab:fig11} b, f, j),
$t=2.496\times10^5 \tg$ (Figure \ref{lab:fig11} c, g, k)
and $t=2.508\times10^5 \tg$ (Figure \ref{lab:fig11} d, h, l).
Three shocks appear at $t=2.480\times10^5 \tg$ (b, f, j):
but the outer shock moves inward
at $t=2.480\times10^5 \tg$, while the inner shocks tend to collide, and ultimately one shock survives
at $t=2.508\times10^5 \tg$. The shock locations are marked by downward arrows for two epochs $t=2.474\times10^5 \tg$
and $t=2.480\times10^5 \tg$.
This pattern occurs repeatedly.
The jet off state ($t=2.508\times 10^5\tg$) is clearly seen in Figure\ref{lab:fig10} (f), where the bipolar outflow perishes.
All the snapshots of Figures \ref{lab:fig10} and \ref{lab:fig11} are from one episode of an oscillating shock starting from a high jet state to
its declining state, are shown in Figures \ref{lab:fig12} (a)--(d), by two dashed vertical lines. Note that the episodic jet ejections
do not constitute relativistic
ballistic ejections but rather these ejections result in continuous stream of jet blobs which constitutes a quasi-steady jet.
In order to quantify the mass outflow rate, we define
\begin{equation}
{\dot m}_{\rm out}=\int \rho v_{\rm out} dA(\mbox{ outer edge})
\end{equation}
and
\begin{equation}
 {\dot m}_{\rm inj}=\int \rho v_{\rm inj} dA(\mbox{ upto }\hinj),
\end{equation}
where $dA$ is the elemental surface area. The matter which is flowing with $v_z>0$ and $v_r>0$
at the outer edge of the computational box is considered as a jet. The relative outflow rate is
\begin{equation}
R_{\dot m}={\dot m}_{\rm out}/{\dot m}_{\rm inj}.
\end{equation}
To see a simplified case of emissivity of these systems, we estimate the bremsstrahlung emission
from the flow. The bremsstrahlung emissivity is
$e_{\rm Brem}\propto \rho^2T^{1/2}\propto \rho^2 c_s$ (energy/volume/time). Therefore, the bremsstrahlung loss
through each volume element, apart from constants and geometrical factors, is
$\delta \epsilon_{\rm Br} \propto  e_{\rm Brem} r^2 dr$.
If the radiation is locally isotropic, \ie equal fluxes in the three directions then, 1/3 of $\delta \epsilon_{\rm Br}$
escapes through the top surface (along $z$). One may be tempted to compare this with the factor
of 1/2 associated with energy loss from a Shakura-Sunyaev disk (SSD)! SSD is an optically thick, geometrically thin
disk with negligibly small advection. The viscous energy dissipated is converted into radiation which will be thermalized because the disk is optically thick. Since the optically thick SSD is geometrically thin,
the entire amount of radiation generated has to escape through the top and the bottom surface, which brings in the factor of 1/2. On the contrary, an advective disk like the one simulated here, is neither optically thick nor geometrically thin, $\ie$ $H/r \lsim 1$. Therefore, radiation will advect along $r$ and $\theta$ directions as well as escape along $z$.
Therefore, in absence of proper radiative transfer treatment, we assume only a third of the radiations generated, escapes
along $z$, from the top half the disk. Due to the up-down symmetry assumed, the same is supposed to occur below the equatorial plane.

Then, the intensity ($I_0$) at each grid point is obtained by dividing $\delta \epsilon_{\rm Br}/3$ by the top surface
area of each volume.
The special relativity implies the radiative intensity in the observer frame will be $I=I_0/[\Gamma (1-v_z)]^4$, where the $I_0$ is the intensity in the comoving frame, and $\Gamma$ is the bulk
Lorentz factor.
This transformation is obtained by starting from the first principle that the phase space density of photons
is Lorentz invariant and has been shown by many authors \citep{hs76, mm84, kfm98}.
Moreover, depending on the location of the source of radiation from which the radiation is emitted,
a factor of ${\cal G}$ is to be taken into account to obtain the amount of the
radiation eaten up by the black hole \citep{st83,vkmc15}, where

\begin{equation}
{\cal G}=\frac{\pi - sin^{-1}[3\sqrt3(1-1/R)/(2R)]}{\pi}; ~~R=\sqrt{(r^2+z^2)}.
\end{equation}
All these corrections are included in estimating the bremsstrahlung loss
$\epsilon_{\rm Br}$ at each time step. As the disc become unstable, the radiation emitted
by the flow should exhibit the same fluctuation. While calculating $\epsilon_{\rm Br}$, we express $e_{\rm Brem}$ in units of $e_{\rm Brem}$ at $\rinj$ to make the estimate bremsstrahlung loss dimensionless.

Figure \ref{lab:fig12} (a) shows $\rsh$ with time for $\alpha=0.01$, and Figure \ref{lab:fig12} (b) shows
$R_{\dot m}$ with time. In Figure \ref{lab:fig12} (c), we plot the estimated bremsstrahlung loss $\epsilon_{\rm Br}$ integrated up to $H_{\rm inj}$, while in Figure \ref{lab:fig12} (d) we plot the shock speed in the black hole rest frame with time.
Figures \ref{lab:fig10} and \ref{lab:fig11} correspond to various time
snaps within the marked region of  Figure \ref{lab:fig12} (a)--(d). The mass outflow rate is episodic;
as the shock generally expands from a minima, the PSD loses its upward thrust, reducing $R_{\dot m}$. As $\rsh$
moves inwards, it squeezes more matter out and $R_{\dot m}$ increases. We also notice the occurrence of
intermittent inner shocks in Figure \ref{lab:fig12} (a) as well.
These secondary shocks are not predicted analytically, but they are only witnessed numerically.
It is instructive to note that the radiative loss follows a time series pattern which has an oscillatory
period similar to that of the oscillating shock.
The shock speed versus time plot shows that the shock speed is generally an order of magnitude smaller
than the local sound speed and the
dynamical speed in the post shock flow.
It is to be remembered that viscosity causes the angular momentum to pile up in the PSD giving rise to extra centrifugal forces across it, and viscous dissipation also increases the thermal energy. Both effects would push the shock front outwards, but as the shock tends to expand, the pressure in the PSD dips, limiting its expansion. Meanwhile, gravity will always attract.
Therefore the delicate force balance between all these interactions
sets the PSD in oscillation. Since the PSD is an extended dynamical fluid body, the oscillation is in general, not a simple harmonic one. The shock front while oscillating extends to within $20-50\rg$ in addition to harboring intermittent inner shocks. One can easily find
some smaller period and amplitude oscillations on the top of the larger variety.
Oscillations of such large fluid
bodies of such a complicated manner broaden the power density spectrum, thus reducing the Quality (Q) factor of the oscillation.

In Figure \ref{lab:fig13} (a), (c) and (e), we plot $\rsh$, $R_{\dot m}$ and $\epsilon_{\rm Br}$,
for $\alpha=0.02$,
and in Figure \ref{lab:fig13} (b), (d) and (f), we plot $\rsh$, $R_{\dot m}$, and $\epsilon_{\rm Br}$, respectively, for $\alpha=0.03$.
As the $\rsh$ oscillation amplitude increases, the secondary shocks get stronger and
the amplitude of $R_{\dot m}$ also increases. Interestingly, there is not only one secondary inner shock
but also are multiple shocks, and the dynamics of these shocks are messy; when an outer shock contracts,
the inner one may expand and collide with the incoming outer shock. $R_{\dot m}$ also increases from
a few percent of the accretion rate to few tens of percent. Since there are many shocks and the
outflowing jet interacts with the surface of the accreting material, the dynamics
of the shocks are also not regular. The bremsstrahlung emission also follows a similar pattern as that of the
shock oscillation.

Figures \ref{lab:fig14} (a), (b) and (c) compare the power spectral density
of the radiation emitted by the accreting fluid which harbors oscillating shocks.
The presence of multiple shocks, their dynamics, as well as the interaction of the outflowing jet and the accreting
matter makes the shock oscillate irregularly, and hence the power spectral density shows multiple peaks.
The outer shock position on an average goes from a maxima to a minima in about $9\times10^3 \tg$ for $\alpha=0.01$,
with many small oscillations on the top of it. The period of these small oscillations is
about $1500\tg$. This gives two frequencies of $0.8$ Hz
and $6.6$ Hz, respectively, if
the central black hole is assumed to be of $10M_{\odot}$. Figure
\ref{lab:fig14} (a) shows the power density spectrum of the radiation with two peaks, as well.
For the case $\alpha=0.02$, the shock oscillates between $10\rg$ and $75\rg$, and $R_{\dot m}$ varies
from a negligible value to about $10\%$ (Figure \ref{lab:fig13} a \& c). When increasing the viscosity to $\alf=0.03$,
the $\rsh$ oscillates from $10\rg$ to about $100\rg$ and the mass outflow rate varies between off-state to more than $20\%$
(Figures \ref{lab:fig13} b \& d). The longer period of shock oscillation for $\alf=0.02$ is around $3\times 10^4 \tg$,
and that for $\alf=0.03$ it is $>3\times 10^4\tg$. Assuming $M_B=10M_\odot$, this results in frequencies $0.3-0.4$Hz
(Figure \ref{lab:fig14} b \& c), respectively. But the power density spectrum of the longer period
for $\alpha=0.02$ and $0.03$ is almost washed out and resembles a broad hump around $0.3-0.4$ Hz. 
For all the three cases shown above, the oscillation of $\rsh$ is reflected more clearly from the estimated radiative loss
corresponding to the harmonics. For $\alpha=0.02$ and $0.03$, the power density spectra of the
estimated radiative loss peaks at $\sim 4$ Hz and $\sim 3 Hz$, respectively. It is to be remembered that
the PDS is presented in arbitrary units.
Smaller periods within a larger period give rise to higher frequencies. It may be noted that, for a low $\alpha$, (\ie $<0.01$)
the median location of the oscillating shock is closer to the horizon, and the period of oscillation is
$<10^4\tg$. So assuming $M_B=10M_\odot$, the period obtained is $\lsim 0.1$ sec and frequency of oscillation is
$\gsim 10$ Hz. To summarize,
increasing $\alf$ causes a larger amplitude but lower frequency shock oscillation
for $\alf <$few$\times10^{-2}$, an oscillation which induces a similar oscillation in the emitted radiation.

\subsubsection{High Viscosity Parameter}

In the literature there have been some multi-dimensional viscous accretion simulations around
black holes which harbor accretion shocks \citep{lmc98,lcscbz08,dcnm14}. As far as we know, all of them were
carried out more or less for low viscosity parameters. With the exception of \citet{lrc11}, most of the simulations were either too hot,
or done in a too small box size. In order to avoid the expensive computation time, simulations
were done for an inner few tens of $\rg$ and the boundary conditions were devised
in a way that shock also forms very close to the horizon.
As a result, when the viscosity parameter was increased to $\alpha \gsim$ few $\times 10^{-3}$,
the shock location escapes out of the computation box, which led to the conclusion that higher $\alpha$
does not support shocks. However, our work showed that as $\alpha$ is increased, the amplitude
of shock oscillation increases until around $\alpha\sim 0.1$ when $\rsh$ goes out of the domain,
while for $\alpha \sim 0.2$, the oscillation amplitude of the shock decreases and is within the
computational domain. To illustrate, we plot $-v_r$ (Figure \ref{lab:fig15} a--d), $c_s$ (Figure \ref{lab:fig15} e--h)
and $l$ (Figure \ref{lab:fig15} i--k) measured along the
equatorial plane, for $\alpha=0.3$ for the accretion model M2. The time slots are
$t=1.272\times 10^5$ (a, e, i), $t=1.276\times 10^5$ (b, f, j), $t=1.294\times
10^5$ (c, g, k) and $t=1.3\times 10^5$ (d, h, l). There are clearly two shocks, where the inner shock moves
very close to the horizon at $t=1.294\times 10^5$.
Higher $\alpha$
ensures more dissipation and therefore
higher $c_s$, or, higher temperature (see Figure \ref{lab:fig15} e--h), which in turn reduces weak multiple inner
shocks, and produces
two predominant shocks, one inner and the other outer. The inner shock is still intermittent but stronger.
More importantly, higher $\alf$ ensures significant angular momentum reduction even in the pre-shock disk
(Figure \ref{lab:fig15} i--k). Since the accretion shock is primarily centrifugal pressure mediated, so
lower $l$ near the horizon, actually brings back the shock into the computational domain. However, hotter PSD with higher
$\alf$ creates
a very strong gradient in $l$ within the PSD. This ensures a large amplitude but a relatively shorter period
($\sim 2800\tg$) oscillation.
As the shock travels to distances $>50\rg$, the sound speed in the immediate
post-shock region is few times lower than the flow close to the horizon (Figure \ref{lab:fig15} e--h).
This causes more efficient angular-momentum
transport in the region closer to the horizon than in the immediate post-shock region, which causes a region
of sharp negative gradient of $l$ \ie $dl/dr<0$ (see Figure \ref{lab:fig15} i). This region of extra centrifugal
pressure within the PSD drives the inner shock. The disk model with higher values of $c_s$ and $\alf$ creates inner
shock, but nonetheless makes the PSD much cleaner than the one for a low $\alf$. Jets are also much stronger, and therefore
jets coming out of PSD are more collimated than those for lower $\alf$. Hotter PSD also causes the
shock front to expand faster and trigger a higher frequency oscillation.
Figure \ref{lab:fig16} (a)--(d) shows the density contours
and velocity vector in the entire computational domain for the same time slots.
These accretion flows form multiple shocks, and at certain times the inner
shock may form at the location near the central object as shown in Figure \ref{lab:fig16} (c).
It is also clear that the jet is well collimated and fast. Comparison of Figures \ref{lab:fig10} (a)--(f)
with Figures \ref{lab:fig16} (a)--(d) shows that 
the jet in Figures \ref{lab:fig16} (a)--(d) flows much closer to the axis. The angular momentum is vastly
reduced due to higher $\alf$ in Figures 16 (a)--(d) making the jet flow closer to
the axis.
We also plot the $c_s$ (Figure \ref{lab:fig17} a), $v_z$ (Figure \ref{lab:fig17} b) and $\rho$
(Figure \ref{lab:fig17} c) with respect to $z$ along the first cell in $r$ ($\equiv$ a distance of $0.5 \rg$
from the axis of symmetry), the snapshot of the jet is at $t=1.198\times 10^5\tg$.
The velocity profile shows that close to the axis, matter is blown out as jet (\ie $v_z>0$) from around a height of $30\rg$.
The sound speed ($c_s$) decreases with height, while velocity increases making the jet
supersonic and eventually
undergoes a series of shocks. The jet speed is fairly high ($\sim 0.2c$) 
especially when the distance is $\sim 200\rg$ which is not a distance at which one expects
a jet to reach its terminal speed. Interestingly, the jet velocity profile (Figure \ref{lab:fig17} b)
also does not reach an asymptotic value and continues to increase at $z=200\rg$.

In the following, we compare various properties of flows starting with the same injection parameters,
and with two different but high $\alpha$. Figures \ref{lab:fig18} (a) and (b) show
$\rsh$ with time, while in Figures \ref{lab:fig18} (c) and (d) we show the compression ratio $v_-/v_+$, and in Figures \ref{lab:fig18} (e) and (f), $R_{\dot m}$ with respect to time. In Figures \ref{lab:fig18} (g) and (h), we plot the power density spectrum (in arbitrary units) of the radiation emitted by the flow. Figures \ref{lab:fig18} (a), (c), (e), (g), or the left
panels are plotted for viscosity $\alpha=0.2$
and Figures \ref{lab:fig18} (b), (d), (f), (h), or the right panels are plotted for $\alpha=0.3$.
Figures \ref{lab:fig18} (a) and (b) show the median of the oscillating shock that has formed closer to the
central object as $\alf$ is increased from $0.2~\rightarrow~ 0.3$. The compression ratio of the oscillating shock may far exceed the steady state values. However, in the case of $\alpha = 0.3$, the compression ratio is obviously higher because the median of the shock is located closer to the black hole. The corresponding mass outflow rate for $\alf=0.3$ is slightly higher than that for $\alf=0.2$.
If the shock is located closer to the inner zone, then the frequency of oscillation should also be higher. For $\alf=0.3$ the frequency of oscillation is around $4$ Hz, while for $\alf=0.2$ it is $\sim 3.5$ Hz. Although both peaks are broad, the peak for $\alf=0.2$ is comparatively broader. The Quality factor of the peaks in the power density spectra are $\sim 2$ for $\alf=0.2$
and $\sim 3$ for $\alf=0.3$. It is interesting to note that for a viscosity of $\alf\lsim$few$\times10^{-2}$, the shock expands with increasing $\alf$, while for $\alf \sim  $few$\times 0.1$, the trend is the opposite. We will discuss this in the next section.


\section{SUMMARY AND DISCUSSION}

In this paper, we simulated the evolution of advective, viscous accretion disk. But instead of randomly chosen values of
injected flow variables, we adopted the values
from the analytical solutions of \citet{kc13}. Excellent agreement of simulation result when they achieved steady state with the
analytical results, shows that the steady state analytical results are indeed steady,
and that the numerical code is very robust too.
In this paper, we have extended the algorithm of our one-dimensional code \citep{lrc11} to multi-dimension.
We regenerated and compared shocked and shock-free steady state viscous solutions
with those from the earlier theoretical work \citep{kc13}.
We considered a shock-free inviscid solution and a shocked inviscid solution corresponding to two different
boundary conditions (referred to as cases M1 \& M2).
In this paper, we considered both computational settings, M1 and M2, and varied $\alpha$ to obtain steady state as well as time dependent solution. Note that even without any artificial shock conditions given, the shock conditions are inbuilt as in any upwind code, as these codes are based on conservation laws of flow variables which assures sharp reproduction of shocks. Since in each cell all the fluxes are conserved, automatically shocks arise if the preferred conditions prevail in the flow.
Such a shock admits entropy and temperature jump across the shock front. In an ideal
fluid this gives rise to the Rankine Hugoniot jump conditions across the shock front. Such a shock
results in higher entropy, and a higher density post-shock flow, whereas the post-shock flow velocity is smaller. Such hotter, slower, denser regions are susceptible to various dissipative processes
and are radiatively more efficient than the pre-shock flow.

We found that the low angular momentum, shock-free accretion becomes similar to a Bondi flow in the presence of viscosity.
No jet-like flow developed when viscosity was turned on for the shock-free accreting flow with
initial conditions of case M1. However, turning on the viscosity for shocked accretion flow with initial conditions of
the case M2, the shock persists in steady state for lower values of $\alf$, but starts to oscillate at higher $\alf$.
Looking closer, one finds that 
a hotter PSD transports angular momentum more efficiently than the colder pre-shock
disk (see Equations (\ref{mfi.eq}) and (\ref{viscdyn.eq})). As a result, the angular momentum distribution becomes
steeper in the PSD than the pre-shock disk, causing an extra centrifugal force on the shock front to
push it out, but the sum of ram pressure and gas pressure of the outer disc would oppose the expansion.
The net effect is that for small $\alf$, the accretion shock settles down to a steady value.
But above a certain critical
viscosity parameter ($\alfcr$), the shock starts to oscillate, and the mass outflow in the form of bipolar jets
increases in strength. In the particular case of M2, $\alfcr=0.003$. As $\alf > \alfcr$, the shock
initially undergoes small perturbations {but on increasing $\alf$ the shock undergoes small amplitude} regular oscillations.
With even larger $\alf$, the oscillation amplitude
increases, and the oscillation itself becomes irregular. There are multiple factors at hand.
The PSD will expand less towards the incoming pre-shock supersonic flow than in the vertical
direction. In fact, the extra thrust of the oscillating PSD ejects matter in episodes along the vertical direction.
The mass that is being ejected might interact with the infalling matter at the interface which gives rise to
a different kind of perturbation.
Moreover, as $\rsh$ moves out to large distance, the angular momentum transport within the PSD becomes
complicated. The flow near the horizon is much hotter than the flow near the expanding shock front. This causes angular momentum distribution in the PSD to change, from a slow monotonic rise of $l$ peaking at some value when $\rsh$ is small, to, two or more sharp peaks when $\rsh$ is large. This causes multiple shocks to form \citep[see][for details of multiple shocks]{lrc11}.
All of these causes irregular oscillation of shocks. And because of the irregularity,
power density spectrums of the shocks show broader peaks than when the oscillation is more regular
\citep{dcnm14}.

According to \citet{dcnm14}, the mass outflow for small-amplitude regular oscillations is episodic and the period of the episodic mass loss matches with that of the shock oscillation.
Their results also showed the existence of one or few sharp peaks in
the power spectrum of the shock, as well as, of the estimated radiations from the
flow. We also checked the case of ес = 0.005 (Figure 9 b) which also exhibits
regular oscillation, also show sharp fundamental peak ($\gsim~10$ Hz) with higher harmonics somewhat similar to Das et al 2014). Although the fundamental frequency of oscillation was lower for the boundary condition of \citet{dcnm14}, note that \citet{dcnm14} performed a simulation for a comparatively hotter, lower angular momentum flow. In the present case, the flow is colder but of higher angular
momentum. Therefore, apart from the location of the shock, the flow properties
across the shock also affect the QPO frequency.

For irregular large amplitude shock oscillations, we compared the time evolution of mass loss with the shock oscillation, and showed that as the shock front starts to contract, it squeezes more matter in the vertical direction, but as the shock front expands from the $\rsh$ minimal position, the PSD looses the upward thrust and the mass outflow collapses, generating the episodic mass outflow.
We note that there is a significant interval of literally no outflow which corresponds to a jet `off' state. We also confirm that during steady state,
the mass outflow rate from the PSD is either absent or weak. Only when the shock activity becomes intensified and thereby
the PSD oscillates appreciably, the mass outflow rate increases. As the viscosity is increased,
the shock oscillation amplitude increases, which trigger a large amount of mass ejection in the form of jet.
The fundamental oscillation period also increases, and the PSD has a messy structure
with many intermittent secondary shocks. This pattern tends to continue for a disk with $\alpha<0.1$.
For $\alf=0.1$, the oscillation amplitude increases to an extent that it actually exceeds
the computational domain. But interestingly, for $\alf \geq 0.2$, the shock oscillation becomes confined
within the computational box and the frequency of oscillation increases.
 Therefore, our simulation results show that for a lower range of viscosity, i.e ., $\alf\sim$few$\times 10^{-2}$, the median of the oscillating shock
increases with $\alf$, while in the range of $0.1<\alf\lsim$few$\times 0.1$, the median of the shock location decreases
with increasing $\alpha$! The question is why so!

Recently,
\citet{kc13,kc14} had shown for a variety
of equation of states of the accretion disk fluid that $\rsh$ decreases with increasing $\alf$ if the flow starts
from the same outer boundary conditions. The explanation to such behavior is that
a higher $\alf$ causes a higher angular momentum transport, reducing the pre-shock angular momentum of the disc, causing $\rsh$ to shift closer to the horizon. \citet{kc13}, in particular also showed
that the cause of the shock expands with increasing $\alf$ in various simulations \citep[including our previous paper,][]{lrc11},
is the short boundary considered for most simulations. By `short' we do not mean a particular fixed value. Its value actually varies depending on the flow parameters. For some flow parameters,
the angular momentum achieves its local Keplerian value at a distance of few$\times100\rg$, while for others, $l=l_{\rm K}$ is achieved at a distance of $\sim10^5\rg$. Therefore a computational box of few$\times100\rg$ is adequate for the former case, but will be considered `short' for the latter case
\citep[see, e. g., Figures 5 d, e of][]{kc13}.

Viscosity is more effective for a hotter and slower flow as seen in Equation (\ref{domegadr.eq}). Hence, viscosity is more effective in PSD than the colder pre-shock disk.
If the outer boundary is short, then $\alf$ cannot significantly affect the flow properties in the pre-shock disk, but efficiently
transports angular momentum in the PSD.
This causes the angular momentum to pile up in the PSD, while in the pre-shock disk $l$ has a low gradient,
and as a result, the shock front
expands in order to negotiate the increased centrifugal force. As we increase $\alf$, more angular momentum will be piled up
in the PSD, but flow properties in the pre-shock disk will largely remain unaffected, and the shock would expand further.
This is roughly what is expected for lower $\alf$ as shown in our simulations. Moreover, as the shock becomes oscillatory, for similar reason,
both the median shock location and the oscillation amplitude increase with increasing $\alf$. This also causes
the emitted radiation to oscillate with decreasing frequency when $\alf$ is increased.
Why is this trend reversed for higher $\alf$ (e. g., Figure \ref{lab:fig18})?

\begin{figure}


\plotone{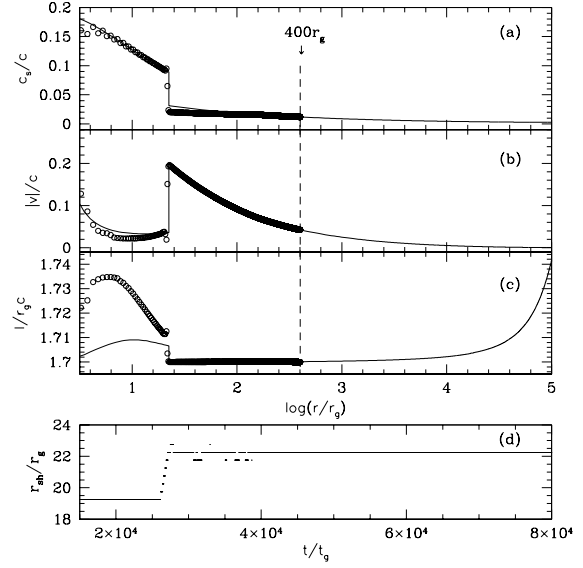}
\caption{Comparison of the theoretical, vertical equilibrium model (solid),
and the two dimensional simulation results on the equatorial plane (open circle) of viscous flow.
The computation box is $200\rg \times 400\rg$ in the $r-z$ plane. The analytical solution is
plotted up to $10^5 \rg$. The injected parameters are
$\vinj=-4.249299\times 10^{-2}~c$, $\linj=1.7\rg c$,
and $\csinj= 1.190908\times 10^{-2}c$ at $\rinj=400 \rg$. The flow variables
are $c_s$ in (a), $|v_r|$ in (b), and $l$ in (c). Locus of shock $\rsh$ with time (d), shows
that $\rsh$ reaches steady state after $t \gsim 4 \times
10^4 \tg$. The viscosity parameter is $\alpha=0.002$. The vertical dashed line denotes the $\rinj$.}
\label{lab:fig7}
\end{figure}

\begin{figure}


\plotone{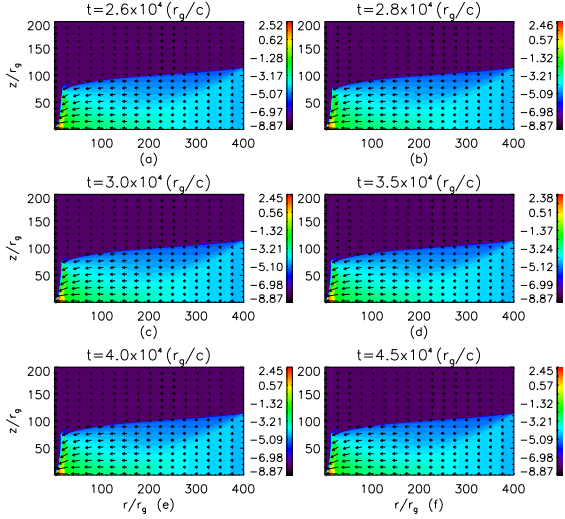}
 \caption{Density contours and velocity fields
of a shocked viscous disk for six time snapshots mentioned on each panel.
The initial conditions are the same as in Figure 7.}
\label{lab:fig8}
\end{figure}

\begin{figure}

\plotone{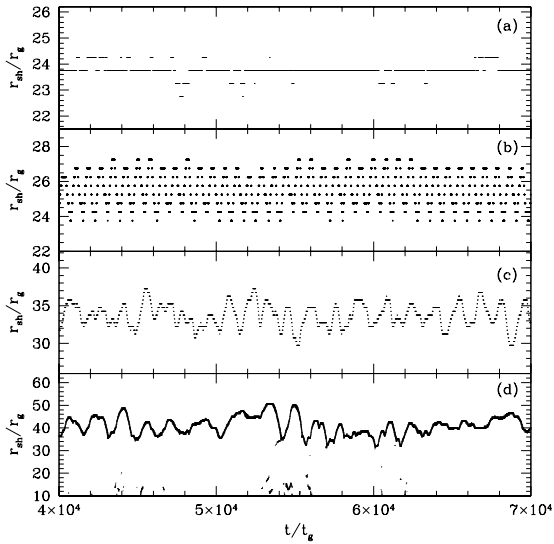}
\caption{Shock location $\rsh$ with $t$. Each panel
is for different viscosity, where (a) $\alpha=0.003$, (b) $\alpha=0.005$, (c) $\alpha=0.007$,
and (d) $\alpha=0.01$. The boundary conditions are same as M2 and the initial condition is the steady state of M2.}
\label{lab:fig9}
\end{figure}

The computational box of $400\rg$, though larger
than most simulation set-ups, is still much smaller compared to the actual size of the theoretical accretion disk
(see Figures \ref{lab:fig7} a--c for comparison).
To understand the situation, let us first focus on Figure \ref{lab:fig7}, where we compared the steady state numerical solution
with the analytical one for the same values of
$\vinj,~\csinj,~\linj$
at $\rinj$. It is clear $\rinj$ is not the
actual outer boundary ($10^{5}\rg$). For a low $\alf$, the angular momentum at the outer boundary will be $l|_{r=10^5\rg} \gsim \linj$.
As we increase $\alf$,
for the same injected values at the same $\rinj$, $l$ at $r=10^5\rg$ will be larger
and for some value of $\alf=\alf_{\rm k}$, the $l$ will attain its Keplerian ($l_{\rm K}=r^{3/2}/[{\sqrt{2}}(r-1)]$)
value at $10^5\rg$. Then for any $\alf>\alf_{\rm k}$, the $l$ distribution will attain its Keplerian value at distance shorter than
$10^5\rg$. Note that for advective-transonic disks, the boundary at which the disk attains $l=l_{\rm K}$,
has to be the maximum value of its outer boundary.
For $\alf \lsim$ few$\times 0.01$, \ie small $\alf$, $l$ does not attain $l_{\rm K}$ within $10^5\rg$.
But for $\alf \sim$ few$\times 0.1$ the outer boundary effectively comes closer, simply because we have kept
the injection parameters constant. By the same token,
$l$ will be substantially reduced as we go inward from $\rinj$ up to the $\rsh$ \citep[for details, see Figure 5 of][]{kc13},
causing the shock position to relocate closer to the central object.
So although $\rinj= 400~\rg$ is still properly not the outer boundary, for $\alf \sim$few$\times 0.1$, the same $\rinj$ is closer to the outer boundary, therefore `mimicking' the fact that with the increase of $\alf$, the shock moves closer to the central object.
Meanwhile, for $\alf \lsim$few$\times 0.01$, $\rinj$ is nowhere close to the real outer boundary. This is the reason why we see $\rsh$ increased with $\alf$ for the range of a $\alf <$few$\times 10^{-2}$, but $\rsh$ decreased with increasing $\alf$ for $\alf \gsim$few$\times 0.1$. The bottom line is that in simulation boxes with a short boundary, we are actually comparing
accretion flows with different outer boundary conditions, where incidentally for a small range of higher $\alf$, $\rinj$ somewhat mimics the outer boundary.

\begin{figure}


\plotone{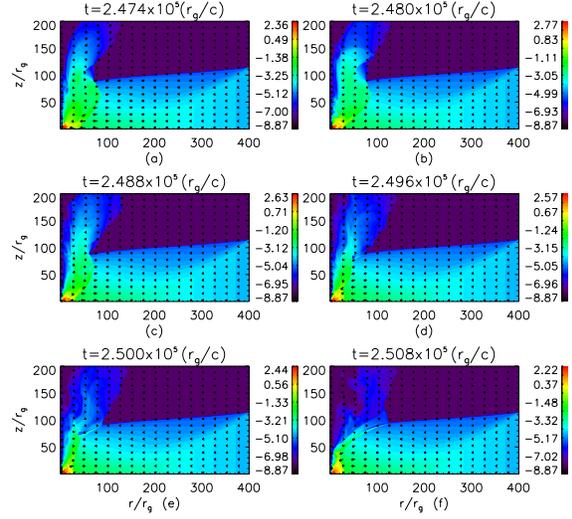}
\caption{Snapshots of density contours and velocity fields in the $r-z$ plane
for $\alpha=0.01$. The initial condition is the steady state of M2 and the boundary conditions
are same as that of M2.}
\label{lab:fig10}
\end{figure}

\begin{figure}

\plotone{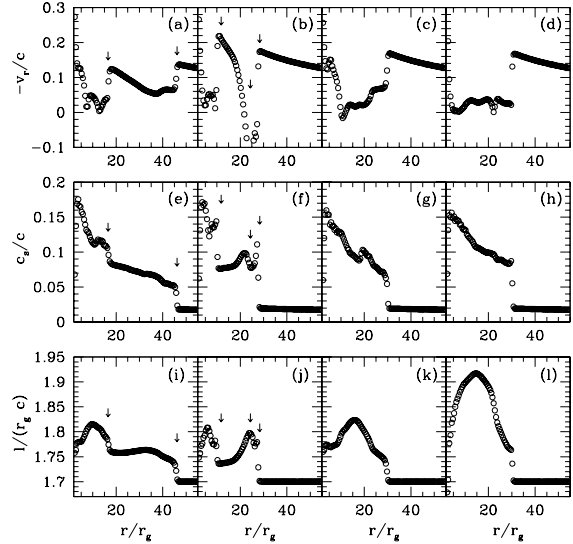}
\caption{Snapshots of $-v_{\rm r}/c$ (a, b, c, d), $c_s/c$ (e, f, g, h) and $l/(\rg c)$ (i, j, k, l), measured on the
equatorial plane, at $t=2.474\times10^5 \tg$ (a, e, i), $t=2.480\times10^5 \tg$ (b, f, j),
$t=2.496\times10^5 \tg$ (c, g, k)
and $t=2.508\times10^5 \tg$ (d, h, l). The downward arrows show the location of shocks.
The slides indicate the same time snaps as in Figure \ref{lab:fig10}.}
\label{lab:fig11}
\end{figure}

\begin{figure}


\plotone{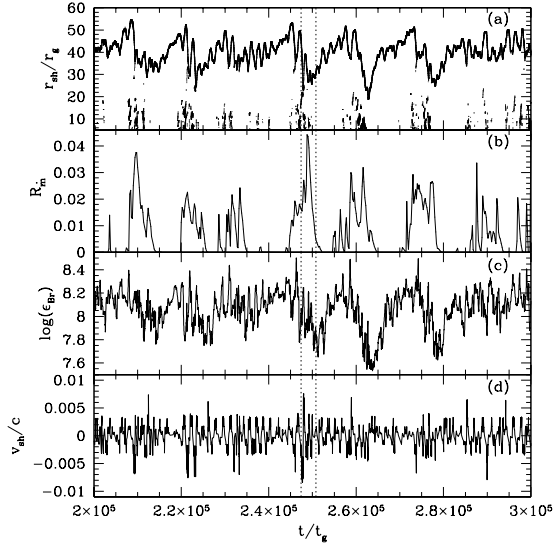}
\caption{(a) Variation of $\rsh$ with time. (b) Variation of $R_{\dot m}$ with time. (c) Variation of $\epsilon_{\rm Br}$ bremsstrahlung emission with time. And (d) shock speed as a function of time.
The viscosity is $\alpha=0.01$, and the snapshots in Figures \ref{lab:fig10} and \ref{lab:fig11} are from various
times in the rising jet phase depicted within the dotted vertical line.}
\label{lab:fig12}
\end{figure}

\begin{figure}

\plotone{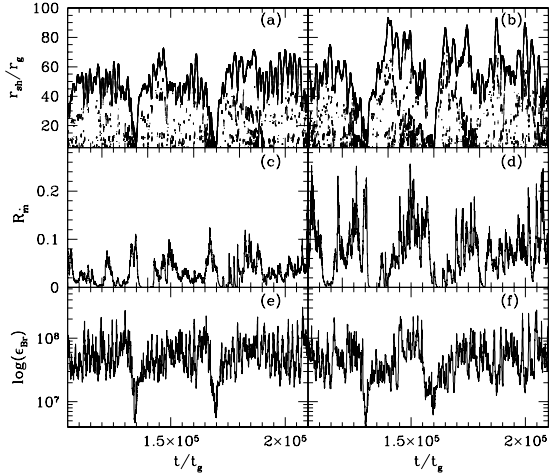}

\caption{Variation of $\rsh$ (a, b), $R_{\dot m}$ (c, d) and $\epsilon_{\rm Br}$ (e, f) with respect to time for
$\alpha=0.02$ (a, c, e) and
$\alpha=0.03$ (b, d, f). The initial condition is the steady state of M2, and boundary conditions are same as those of M2.}
\label{lab:fig13}
\end{figure}

\begin{figure}

\plotone{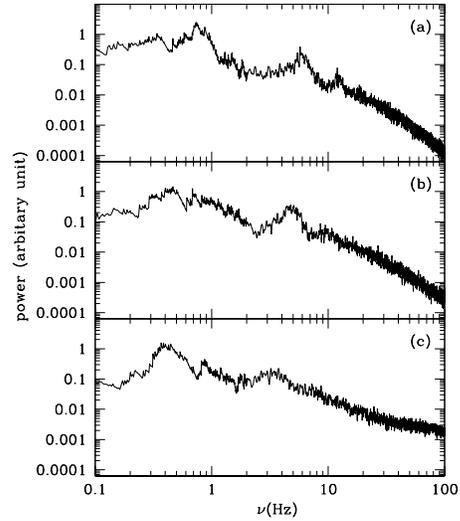}

\caption{Comparison of the power spectral density (arbitrary units) transform of the shock oscillation for $\alpha=0.01$ (a), $\alpha=0.02$ (b),
and $\alpha=0.03$ (c). The spectral density is done considering a stellar mass BH of $M_{BH}=10M_\odot$.}
\label{lab:fig14}
\end{figure}

\begin{figure}
\plotone{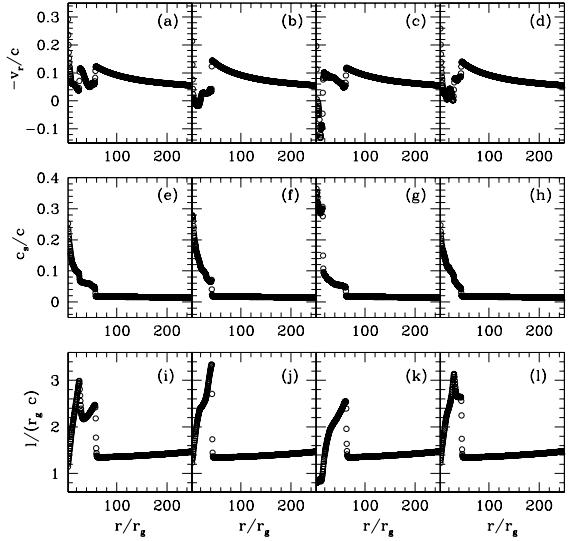}

\caption{Snapshots of $-v_r$ (a, b, c, d), $c_s$ (e, f, g, h) and $l$ (i, j, k, l) measured
in geometrical units, and evaluated on the equatorial plane. Various time snaps
are at $t=1.272\times 10^5$ (a, e, i), $t=1.276\times 10^5$ (b, f, j), $t=1.294\times
10^5$ (c, g, k), and $t=1.3\times 10^5$ (d, h, l). The viscosity parameter $\alpha=0.3$ and initial condition is the
steady state of M2 and the boundary conditions are also identical to those of M2.}
\label{lab:fig15}
\end{figure}

\begin{figure}

\plotone{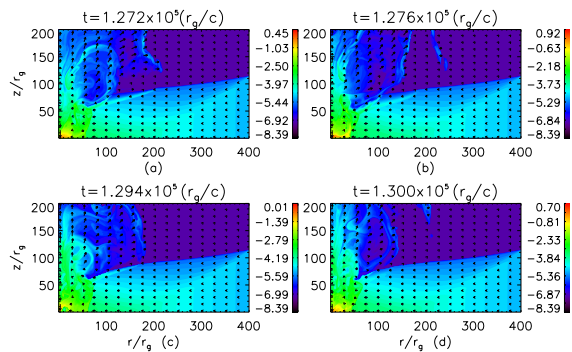}

\caption{Contour of density and velocity vectors of an accretion disk and its jet. Various time snaps
are $t=1.272\times 10^5$, $1.276\times 10^5$, $1.294\times
10^5$, and $1.3\times 10^5$. The viscosity parameter is $\alpha=0.3$, and is the complete solution of
Figure \ref{lab:fig15}.}
\label{lab:fig16}
\end{figure}

\begin{figure}

\plotone{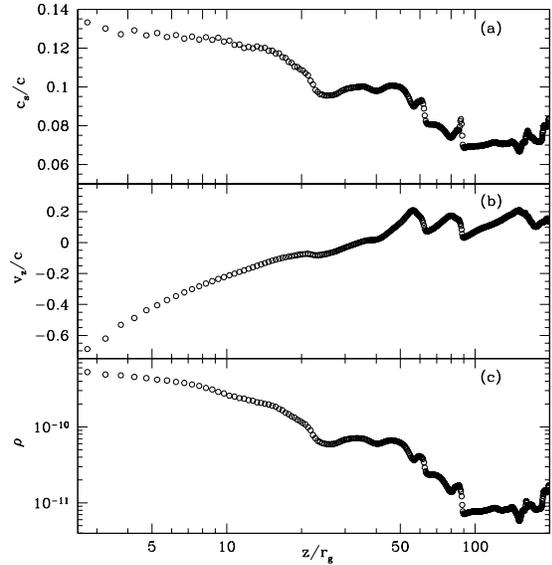}

\caption{Jet profile plotted along $z$ coordinate during the `jet on state' at $t=1.198\times10^5\tg$
same case as
Figure \ref{lab:fig15}. The sound speed (a), $v_z$ (b) and density $\rho$ (c).
The flow variables plotted are taken form the first cell adjacent to the axis.}
\label{lab:fig17}
\end{figure}

\begin{figure}

\plotone{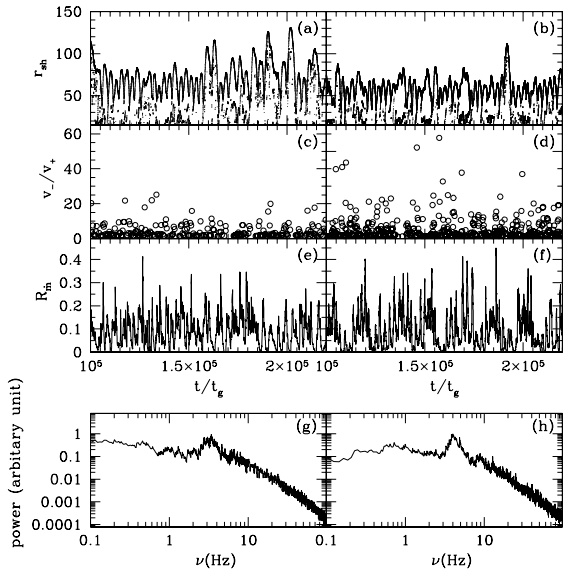}

\caption{Variation of $\rsh$ (a, b), compression ratio $v_-/v_+$ (c, d), $R_{\dot m}$ (e, f) with time. The power density spectrum
in arbitrary units (g, h) for the two viscosity cases are plotted with frequency.
The left panels (a, c, e, g) represents flow with $\alf=0.2$ and the right
panels (b, d, f, h) represents flow with $\alf=0.3$. The boundary condition is same as that of M2.}
\label{lab:fig18}
\end{figure}

The mass outflow rate for higher $\alf$ appears to be sporadic, with an inconspicuous jet `off' states.
Since the viscosity is very strong for $\alf=0.3$, a higher viscous dissipation and more significant angular momentum
transport induce a higher frequency shock oscillation. The jet becomes much stronger
at $\alf=0.3$,
to the extent that average jet speed near the axis
is $\sim 0.2c$ at a height of $200\rg$ above the equatorial plane. One may wonder whether we should call these
outflows as jets, given the fact that these are not truly relativistic. We note two points in the jet characteristics. First, jets are collimated
ejections. Figures \ref{lab:fig10} and \ref{lab:fig16} clearly show that the outflow is fairly collimated (the bulk of it is spread within $100\rg$ at a height of $200\rg$).
Next, these outflows leave the computational domain at $v\sim 0.2c$, which is mildly relativistic and
clearly transonic (Figures \ref{lab:fig17} a \& b). So according to these conditions, they qualify as jets.
From Figure \ref{lab:fig17} (b), the jet is obviously not reaching its asymptotic
value at the height of $200\rg$; therefore a somewhat higher speed can
be expected at $z > 200\rg$.
However, this is not an indication that this jet will
go on to reach a relativistic terminal speed. One must also bear in mind that
not all jets, especially those around micro-quasars, are always truly relativistic
(S433 Margon 1984, and 2009 burst of H1743-22 Miller-Jones et al. 2012).
Our simulation set-up does not address the transition from intermediate states to the high soft state (or, transitions across the jet line) and the associated ejection of relativistic blobs. We simulate the origin of semi relativistic jets associated with the low hard state and the intermediate states. And indeed such jets increases in strength as the black hole candidates move from low hard to intermediate hard spectral states \citep{fbg04}.

In various papers, many authors have shown
that in out-bursting sources the low QPO frequencies emerge in the hard states and increases
as the object transits from low hard states to the intermediate states. Such a QPO is not detected during the ejection of relativistic jets \citep{cbhs04,mc06,ndmc12}. In the model, the shock being situated at large distances is equivalent to a low hard state, and as the median of the oscillating shock
moves towards the central object, the total disk luminosity increases.
Any perturbation of the shock, while as a whole moving towards the central object, would increase the frequency
of the oscillation. Simultaneously, the mildly relativistic jet becomes stronger \citep[Q diagram of][]{fbg04}, as also seen in our
simulation. Although we could not track its entire evolution because of the limitation of the simulation box size, at least
for higher $\alf$, the increase of the QPO frequency and strengthening of the mildly relativistic jets somewhat justify the theoretical conjecture \citep{kc13,kc14,ck16}.
However, the
whole set of state transition can be emerged if and only if one simulate an accretion flow from the actual outer boundary
(where $l = l_{\rm K}$ , or, $\rinj = 10^5 \rg$)
and higher $\alf$, which is very challenging to achieve and presently beyond the scope of this paper.

There have been other interesting investigations in the advective flow regime,
for instance, general relativistic hydrodynamic simulations \citep{ny09}
and investigations of transmagnetosonic flow in general relativity
\citep{trft02,tgfrt06,ftt07}. While \citet{ny09} only simulated inviscid flow and reported a
shock oscillation of few Hz, the main importance is that it was possible to obtain
steady and oscillatory shocks in general relativistic simulations. The transmagnetosonic flow also reported the formation of general relativistic MHD shocks.
The presence of both slow MHD shocks and fast MHD shocks opens up hitherto
uncharted possibilities. Fast shocks may generate transverse magnetic fields,
which can help in powering jets. An interesting investigation may be taken up
to identify various spectral states with the type of MHD shocks.

Presently, we conclude that shocked accretion disk through the oscillation of PSD
naturally explains the QPO phenomena in black hole candidates, while episodic
jet seems to get stronger as viscosity increases. For weak viscosity the jet is also
weaker, while an oscillating shock due to its `bellow action', is squeezing out episodic
jets at fairly high speed. In \citet{lrc11}, the median shock location was large
and therefore the frequency of oscillation obtained was around $0.1$ Hz, whereas in \citet{dcnm14}, the median
shock location was at few$\times 10~\rg$. In addition, the frequency was around a few Hz. In this paper,
we investigated a large range of viscosity parameters but starting with the same initial condition,
and we were able to generate frequency ranges from less than one to few Hz. Moreover,
\citet{lrc11}, being an one-dimensional analysis, failed to simulate shock oscillation beyond
$\alf >0.1$, but following the conjecture by \citet{lrc11}, we show that formation of jet/outflows in multi-dimensional
simulations saturates the shock oscillation
for higher $\alf$ and keeps the shock oscillation within the computational domain, where it is shown that transient shock
survives even in high viscosity parameters, and the mass outflow rate also becomes stronger for such a flow.

\acknowledgments{SJL and SH would like to acknowledge from the Basic Science Research Program
through the National Research Foundation of Korea (NRF 2015R1D1A3A01019370; NRF 2014R1A1A4A01006509).
DR was supported by the National Research Foundation of Korea through grant 2007-0093860 and 2016R1A5A1013277.}

\end{document}